\documentclass[aps,prb,twocolumn,amssymb]{revtex4}

\usepackage{graphicx}
\usepackage{graphics}
\usepackage{dcolumn}
\usepackage{bm}
\usepackage{amsmath}
\usepackage{color}
\usepackage{dsfont}
\usepackage{tikz}
\usetikzlibrary{decorations.pathreplacing} 
\usepackage{xfrac}

\begin{document}

\newcommand{\bk}{{\bf k}}
\newcommand{\bp}{{\bf p}}
\newcommand{\bv}{{\bf v}}
\newcommand{\bq}{{\bf q}}
\newcommand{\tbq}{\tilde{\bf q}}
\newcommand{\tq}{\tilde{q}}
\newcommand{\bQ}{{\bf Q}}
\newcommand{\br}{{\bf r}}
\newcommand{\bR}{{\bf R}}
\newcommand{\bB}{{\bf B}}
\newcommand{\bA}{{\bf A}}
\newcommand{\ba}{{\bf a}}
\newcommand{\bE}{{\bf E}}
\newcommand{\bj}{{\bf j}}
\newcommand{\bK}{{\bf K}}
\newcommand{\cS}{{\cal S}}
\newcommand{\vd}{{v_\Delta}}
\newcommand{\tr}{{\rm Tr}}
\newcommand{\kslash}{\not\!k}
\newcommand{\qslash}{\not\!q}
\newcommand{\pslash}{\not\!p}
\newcommand{\rslash}{\not\!r}
\newcommand{\bs}{{\bar\sigma}}
\newcommand{\omt}{\tilde{\omega}}

\newcommand{\vv}{\mathcal V}

\newcommand{\qperp}{q_{\perp}}
\newcommand{\qpar}{q_{\parallel}}
\newcommand{\beq}{\begin{equation}}
\newcommand{\eeq}{\end{equation}}

\newcommand{\redtext}[1]{\textcolor{red}{#1}}
\newcommand{\bluetext}[1]{\textcolor{blue}{#1}}
\newcommand{\purple}[1]{\textcolor{purple}{#1}}

\newcommand{\ket}[1]{| #1 \rangle}
\newcommand{\bra}[1]{\langle #1 |}
\newcommand{\dirac}[2]{\langle #1 | #2 \rangle}

\title{Transport through a disordered topological-metal strip}

\author{Alexandra Junck,$^{1,2}$ Kun W. Kim,$^1$ Doron L. Bergman,$^1$ T. Pereg-Barnea,$^{1,3}$ and Gil Refael$^1$}
\affiliation{$^1$Department of Physics,
California Institute of Technology, 1200 E. California Blvd, MC114-36,
Pasadena, CA 91125 }
\affiliation{$^2$Dahlem Center for Complex Quantum Systems and Fachbereich Physik, Freie Universit\"at Berlin, 14195 Berlin, Germany}
\affiliation{$^3$Department of Physics, McGill University, Montreal, Quebec, Canada H3A 2T8}
\date{\today}
\begin{abstract}
Features of a topological phase, and edge states in particular, may be
obscured by overlapping in energy with a trivial conduction band. The
topological nature of such a conductor, however, is revealed in its
transport properties, especially in the presence of disorder. In this
work, we explore the conductance behavior of such a system with
disorder present, and contrast it with the quantized conductance in an ideal 2D
topological insulator. Our analysis relies on numerics on a lattice
system and analytics on a simple toy model. Interestingly, we find that
as disorder is increased from zero, the edge conductivity initially
falls from its quantized value; yet as disorder continues to increase,
the conductivity recovers, and saturates at a value slightly below the
quantized value of the clean system. We discuss how this effect can be
understood from the tendency of the bulk states to localize, while the
edge states remain delocalized.
\end{abstract}
\maketitle

\section{Introduction}

To date, topological behavior has been observed in many systems in zero magnetic field, usually by identifying topologically induced edge states. Much of the  theoretical attention has been given to topological insulators, \cite{KaneMele1,KaneMele2,FuKane3D,FuKaneMele3D,MooreBalents,QiZhangIndex,QiZhangTR} while many of the experimental observations are of systems that are metallic, due to either doping, or midgap states. \cite{Jarillo-Herrero,Paglione,Hsieh,HasanTernery} It is natural then to ask - what kind of topological behavior can a metal, or a gapless system, exhibit? 

Several groups have studied the so-called topological conductor. Ref.~\onlinecite{BergmanRefael} used a 2d Kane-Mele (or Haldane) topological insulator, with interstitial sites in each hexagon forming an additional trivial band to realize a topological conductor (see Fig.~\ref{fig:model_conventions} for illustration). The 'parasitic' metallic band was made to overlap with the topological-band edge states. Hybridization between the topological and parasitic metallic bands changed the edge-state spectrum in a peculiar way. Edge states which overlapped in both energy and parallel momentum with bulk metallic states not only did not disappear when hybridization was introduced, but rather, they doubled: Exact edge states emerged at energies above and below the metallic band, and a finite width spectral resonance remained where the unhybridized topological edge state used to be. An additional topological conductor system was studied by Bergman\cite{Bergman} and Barkeshli and Qi.\cite{BarkeshliQi} They showed that a magneto-electric axion response (with time-reversal broken at the surface) persists even when the 3d topological insulator is doped, and Fermi-surfaces appear in its bulk, although it is no longer quantized. Similarly, Ref.~\onlinecite{Plasmon} showed that a 3d topological metal still supports a special surface plasmon mode. 

The transport properties of a topological conductor, especially in the presence of disorder are the focus of this paper. These aspects of the topological conductor were so far mostly ignored. Several groups, however, have investigated the effects of disorder on topological insulators. It was clearly shown that a sufficient amount of disorder will close a topological gap, resulting in a metal. \cite{Shindou1,Shindou2,Prodan2D,Prodan3D,HastingsLoring1,HastingsLoring2,Ryu,Voijta,MeyerRefael} Furthermore, it was found that disorder could even induce topological behavior in trivial semiconductors with spin-orbit coupling.\cite{Jain,Beenakker,GuoFranzRefael} Only Ref.~\onlinecite{BergmanRefael2} so far also considered disorder effects in a topological-metal regime, and found that it does not qualitatively affect the non-universal magneto-electric effects found by Refs.~\onlinecite{BarkeshliQi,Bergman}. 

Our study concentrates on the 'parasitic metal' flavor of the topological conductor which was described above.\cite{BergmanRefael} Specifically, we investigate the conductance of a topological conductor strip as a function of disorder, with its chemical potential in the energy range where edge states and the parasitic metallic band coexist. Increasing disorder will eventually localize the bulk states, and then we expect there to be rather little mixing between the bulk states and the edge states, since the wavefunction overlaps are exponentially small. In contrast, for weak disorder, where the bulk states are still mostly delocalized, the wavefunction overlaps are much more significant, and we expect that no state will be distinguishable as an edge state. 

Our results, however, are quite surprising. The edge states appear to retain their significance for essentially all disorder strengths. Their presence and distinction is manifested in the persistence of a single highly conducting channel which survives until the disorder is sufficiently strong and the topological band associated with the honeycomb subsystem is destroyed. The distinct feature of this effect is the appearance of a conductance minimum for the most conducting channel at some finite disorder (see Fig.~\ref{schematicdip}). We characterize this conductance minimum in systems of varying width and length, as well as for several values of the model parameters.

The organization of the paper is as follows. In Sec.~\ref{model} we describe in detail the parasitic metal model we study. We describe the transfer-matrix method we use in Sec.~\ref{methods}. Our results, demonstrating the conductance minimum, as well as its dependence on the system parameters are recounted in Sec.~\ref{results}. A qualitative understanding of the effect can be obtained using a simple model which we describe and analyze in Sec.~\ref{analytics}, before making our concluding remarks in Sec.~\ref{conc}.

\section{The Kane-Mele parasitic band model for topological conductor \label{model}}

We perform the disorder and transport analysis on a specific model for the topological conductor. We expect that the qualitative behavior will be independent of the specific model we employ. As mentioned in the introduction, we will use the Haldane model with interstitial sites added. We will also consider the full Kane-Mele model with Rashba interaction included and find a similar behavior as for the Haldane model.

 \begin{figure}
 	\centering
 		\includegraphics[width=1.8in]{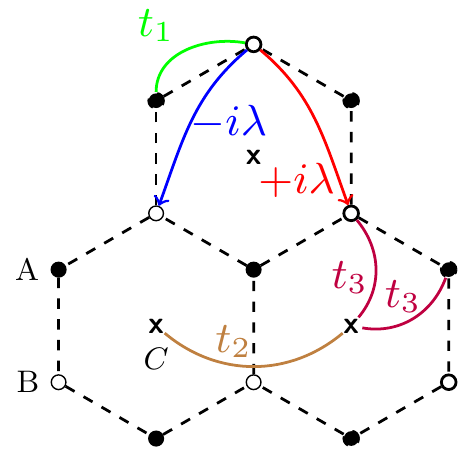}
 	\caption{A two dimensional lattice model of a topological conductor. The model is based on the Kane-Mele 
	model\cite{KaneMele1,KaneMele2} on the honeycomb lattice, which is denoted by dashed lines. The Kane-Mele model
	consists of nearest-neighbor hopping $t_1$ (green) between the two ($A,B$) sublattices of the honeycomb lattice 
	(denoted by empty and filled circles respectively), and a complex second-nearest-neighbor hopping (a spin orbit coupling term)
	with opposite sign when clockwise ($+i \lambda$, red) and counterclockwise ($-i \lambda$, blue),
	as indicated by the curved arrows in the figure.
	In addition to sites of the honeycomb lattice, we include a new set of sites ($C$) at the centers 
	of the hexagonal plaquettes of the honeycomb lattice, denoted by cross marks. The $C$-sites form a triangular sublattice, 
	and nearest-neighbor hopping between them $t_2$ (brown) forms a metallic band. 
	To explore the interplay between the topological insulators helical surface states and the bulk metallic band we mix the 
	two systems by allowing hopping between the $C$-sites and the honeycomb lattice sites, $t_3$ (purple). 
 	}
 	\label{fig:model_conventions}
 \end{figure}

The Haldane model\cite{Haldane} consists of spinless electrons on the honeycomb lattice, with nearest-neighbor hopping which gives the band structure of Graphene, and imaginary second-nearest-neighbor hopping which opens a gap, induces a non-zero integer Chern number in the conduction and valence bands, and produces chiral edge states. To this we add a parasitic metallic band, formed by an overlayed triangular lattice of sites, which occupy the center of the honeycomb hexagons (see Fig.~\ref{fig:model_conventions}). Hopping between the interstitial triangular lattice sites produces a single topologically-trivial band, which overlaps in energy with the edge states of the Haldane model. We then allow the two subsystems to hybridize. The Hamiltonian describing the combined model is
\begin{eqnarray}\label{Hamiltonian}
{\cal H} &=& {\cal H}_{Haldane}+{\cal H}_{met}+{\cal H}_{hyb} + {\cal H}_{dis} \nonumber \\
{\cal H}_{Haldane} &=& - t_1 \sum_{\langle i j \rangle } a^{\dagger}_{i} b^{\phantom\dagger}_{j} 
- i \lambda \sum_{\langle \langle i j \rangle \rangle}
\left[a^{\dagger}_{i} a^{\phantom\dagger}_{j} \nu_{i j} + \left( a \rightarrow b \right)  \right] +h.c \nonumber \\
{\cal H}_{met}&=&- t_2 \sum_{\langle i j \rangle} \left[ c^{\dagger}_{i} c^{\phantom\dagger}_{j} + h.c. \right] \nonumber \\
{\cal H}_{hyb} &=& - t_3 \sum_{\langle i j \rangle } \left[ c^{\dagger}_{i} a^{\phantom\dagger}_{j} +
c^{\dagger}_{i} b^{\phantom\dagger}_{j} + h.c. \right] \nonumber \\
{\cal H}_{dis} &=& \sum_{i,x \in \{a,b,c\}}V_i x^{\dagger}_{i} x^{\phantom\dagger}_{i}
\; , 
\label{Htotal}
\end{eqnarray}
where $i,j$ denote the sites of the composite honeycomb and
interstitials lattice. The operators $a,b$ denote the fermion
annihilation operators on the two honeycomb sublattices (A and B), and
$c$ is the fermion annihilation operator on the triangular lattice sites $C$ at
the centers of the honeycomb plaquettes. The coefficients $\nu_{i j}=
\pm 1$ determine the sign of the imaginary second-nearest-neighbor
hoppings, and are defined in
Refs.~\onlinecite{KaneMele1,KaneMele2}, as $\nu_{ij}=\frac{2}{\sqrt{3}} {\hat z} \cdot (\hat{d}_{jk} \times \hat{d}_{ki})$, where $\hat{d}_{kj}$ is the unit vector pointing
from site $j$ to $k$, and $k$ is the intermediate site between $j$ and
$i$. The parameter $t_1$ denotes the nearest-neighbor hopping on the honeycomb lattice, while $t_2$ denotes the nearest-neighbor hopping on the triangular lattice.
Finally, $t_3$ is the hybridization hopping between the honeycomb and triangular lattice sites. The random on-site potential $V_i$ on every site of the lattice (honeycomb and triangular lattice sites) has a uniform distribution between $[-W,W]$.

\begin{figure}[t]
\begin{minipage}{4.25cm}
	\centering
	      \includegraphics[width=4.5cm]{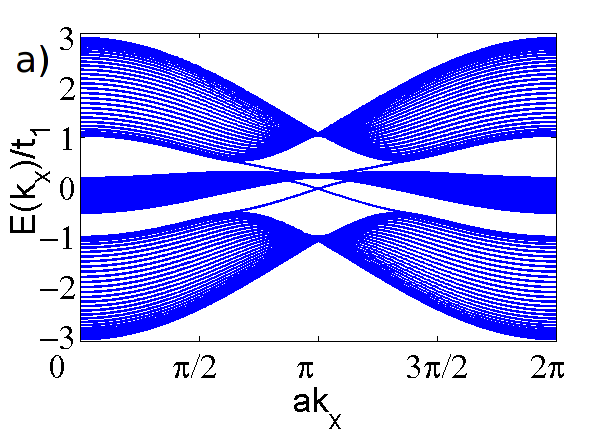}
\end{minipage}
\hfill
\begin{minipage}{4.25cm}
\centering
	    \includegraphics[width=4.3cm]{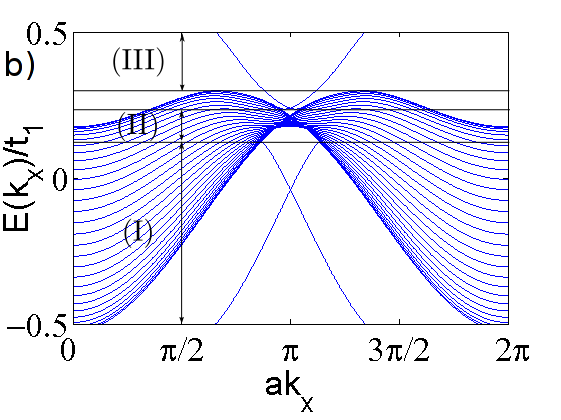}
\end{minipage}
 \caption{(a) Energy spectrum of the topological conductor without disorder. Bands at the top and bottom are the bulk states of the honeycomb lattice. The band in the middle is the metallic bulk from the interstitial sites. The parameters in units of $t_1$ are $\lambda = 0.1 $, $t_2=0.1$, $t_3=0.1$, and the width of the strip is $M=30$ zig-zag lines. (b) Zoom of the band structure in (a). Different Fermi-energy regimes are indicated. The region between (II) and (III) has the same properties as region (I).}
\label{spectrum1}
\end{figure}

The single spin system we analyzed consisted of zig-zag strips of the topological conductor with the Hamiltonian (\ref{Hamiltonian}). Its generic band structure is shown in Fig. \ref{spectrum1}. Indeed we see that the metallic band intervenes between the conduction and valence bands of the Haldane portion of the band structure. Indeed we see regions in the spectrum where in a single value of the momentum parallel to the edge, there are two edge states at energy above and below the metallic band. This is a manifestation of the exiling effect pointed out in Ref. \onlinecite{BergmanRefael}, and indicates that the metallic band appeared on top of the edge-states branch in this region, and hybridization expelled the edge states while doubling them.

\subsection{Rashba coupling}

We have also studied the effect of adding a Rashba type spin-orbit coupling on the honeycomb lattice to our model \eqref{Hamiltonian}, with the electron spin$-\frac{1}{2}$ restored. The Rashba interaction mixes between the two spin flavors as follows
\begin{equation}
{\cal H}_R=i\lambda_R \sum_{\langle ij\rangle \alpha \beta}a^{\dagger}_{i\alpha}(\hat{s}_{\alpha\beta}\times \hat{d}_{ij})_z b_{j\beta} + h.c.,
\label{Hrashba}
\end{equation} 
where $\hat{s}$ is the vector of Pauli matrices for the electron spin, $\hat{d}_{ij}$ is the unit vector pointing from site $j$ to $i$, and $\alpha$, $\beta$ are spin indices.

Even with Rashba interaction included, we find the same qualitative behavior as in the spinless case described above. 

\subsection{Fermi-energy regimes}

 The systems' transport properties depend closely on its Fermi energy. The range of Fermi energies between the conduction and valence bands of the topological subsystem can be split into three important regions. Region I denotes the case where the edge states and metallic bands overlap in energy but not in momentum, and therefore coexist in the clean limit. Region II denotes the energy range of the parasitic metal where no edge states appear. Region III is the energy range that has only edge states. These regions are indicated in Fig.~\ref{spectrum1}(b) for the spinless model.

\section{Landauer formalism for the strip\label{methods}} 

In order to analyze the two-terminal transport through the topological conductor, we use the Landauer-B\"uttiker formalism.\cite{Landauer1957,Landauer1970,Buttiker1988} We envision our system as consisting of a long strip of the topological conductor, which is disordered in a finite region (Fig.~\ref{stripfig}). The regions to the left and right of the disordered region form the ballistic leads,  with several transverse modes. The Landauer-B\"uttiker formalism gives the following formula for the current flowing through the system in a two-terminal device:\cite{Fisher1981}
\begin{equation}
G=\frac{e^2}{h}\sum_{n,m}{|T_{nm}^{LR}(E_F)|^2},
\label{eq:conduc1}
\end{equation}
where $T_{nm}^{LR}(E_F)$ is the transmission coefficient for going from mode $m$ in the left lead to mode $n$ in the right lead at the Fermi energy, $E_F$. To ensure probability current conservation, it is important to note that the current associated with a scattered wave is proportional to the square of the wavefunction multiplied by the velocity. Therefore $T_{nm}$ is given by the outgoing current amplitude, i.e., the wave amplitude times the square root of the velocity of an electron leaving the device in mode $n$ through one lead if the incoming current amplitude in mode $m$ in the other lead is set to unity.                                    

Eq. (\ref{eq:conduc1}) can also be written as the sum of conductances per channel $g_m$:
\begin{equation}
G=\frac{e^2}{h}\sum_{m}{g_m},
\label{eq:conduc3}
\end{equation}
where $g_m=\sum_{n}|T^{LR}_{nm}|^2$ is the conductance of channel $m$, given by the probability that an electron entering the system in mode $m$ is transmitted through the conductor.
We find the transmission coefficients $T_{nm}^{LR}(E_F)$ using the transfer-matrix method, as described in Appendix~\ref{appA}. We note that this method, particularly when disorder is concerned, has an instability which restricts the system size possible. This instability arises due to the existence of imaginary momentum modes at the desired energy range which appear because we are exploring the conductance of the topological conductor for energies in the bulk gap of the honeycomb lattice. Indeed, the number of imaginary momentum modes due to the honeycomb subsystem increases linearly with the number of zig-zag lines in the strip, thus restricting the accessible system size.\footnote{We have $2M-1$ imaginary momentum modes due to the honeycomb system for a strip with $M$ zig-zag lines.} Nevertheless, the transfer-matrix method provides reliable results for a range of strip sizes. These results could, in principle, be improved upon by using S-matrix methods.\cite{Datta}

\section{Results\label{results}}

We investigated the conductance of the topological conductor strip for
different Fermi-energy regimes and analyzed its dependence on system
and model parameters. For Fermi energies where metallic bulk and edge
states coexist (region I), we find a distinct minimum in the
conductance of the most conducting channel as a function of disorder
strength, followed by a revival towards ballistic transport in this
channel. This is a surprising feature given the mixing between bulk and edge.

\subsection{Region I}

\begin{figure}[t]
 \includegraphics[width=8cm]{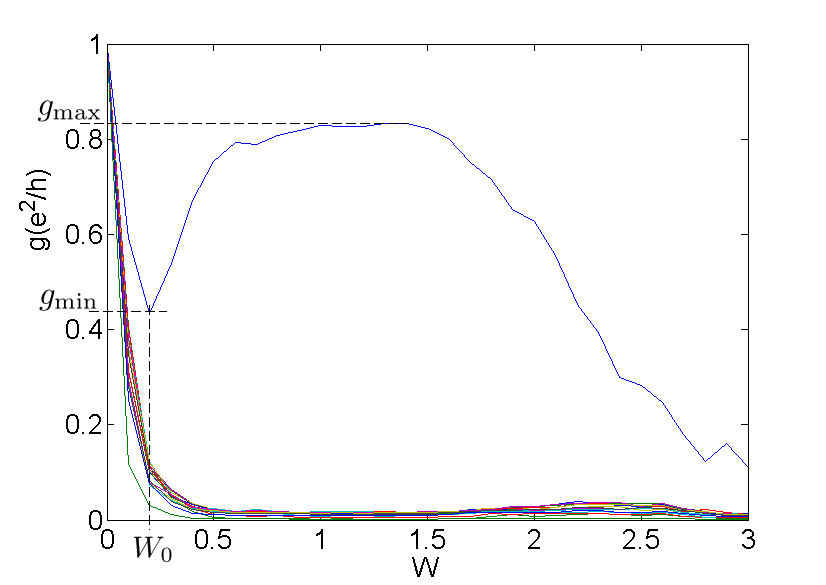}
\caption{Illustration of the 'dip and recovery' feature of the edge channel conductance. Conductance $g_n$ of different channels is plotted as a function of disorder strength $W$. The edge channel conductance shows a sharp minimum $g_{\textnormal{min}}$ at finite disorder $W_0$ and a broad maximum $g_{\textnormal{max}}$. Parameters are $E_F=0.1$, $\lambda=t_3=0.1$, and $M=L=20$.}
 \label{schematicdip}
\end{figure}

The conductance for each channel for a Fermi energy in region I is
shown in Fig.~\ref{schematicdip}. The generic feature we find in this
regime is a single highly conducting
channel that persists up to large disorder and decays when disorder is presumably
strong enough to close the gap of the honeycomb subsystem. For small
disorder there is a sharp dip in the conductance followed by a broad
maximum.

\begin{figure}[t]
\begin{minipage}{4.2cm}
	\centering
		\includegraphics[width=4.5cm]{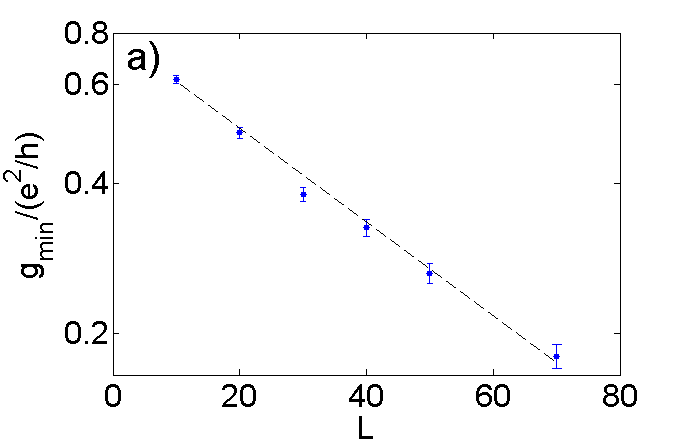}
\end{minipage}
\hfill
\begin{minipage}{4.2cm}
\centering
		\includegraphics[width=4.5cm]{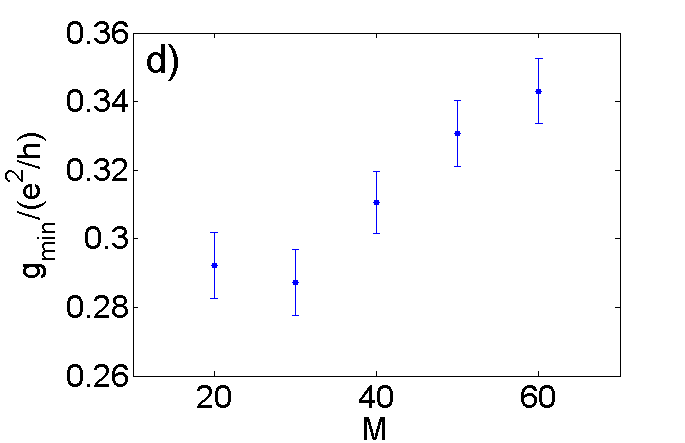}
\end{minipage}
\begin{minipage}{4.2cm}
	\centering
		\includegraphics[width=4.5cm]{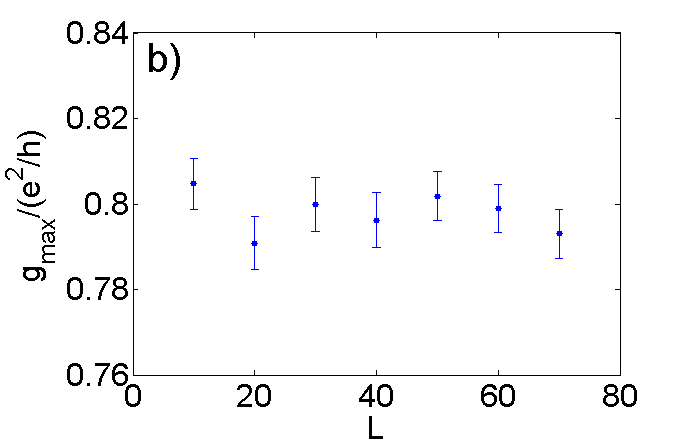}
\end{minipage}
\hfill
\begin{minipage}{4.2cm}
\centering
		\includegraphics[width=4.5cm]{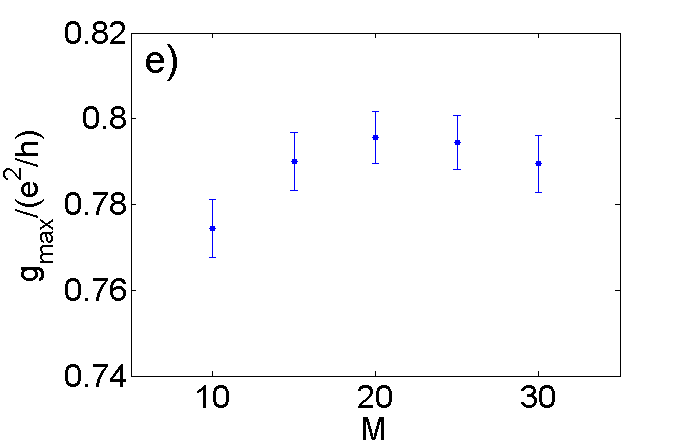}
\end{minipage}
\begin{minipage}{4.2cm}
	\centering
		\includegraphics[width=4.5cm]{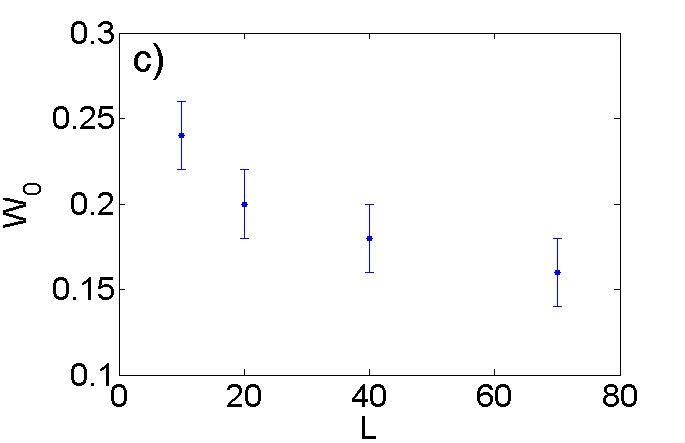}
\end{minipage}
\hfill
\begin{minipage}{4.2cm}
\centering
		\includegraphics[width=4.5cm]{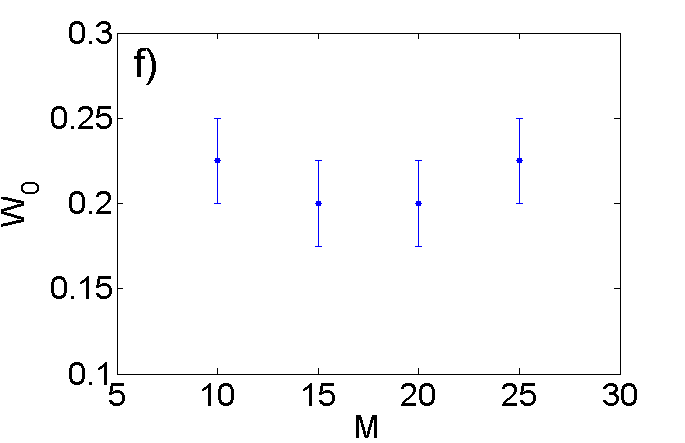}
\end{minipage}
\caption{Dependence of the conductance dip and maximum on system size (Region I). (a)-(c) Dependence on length $L$ of the system for $M=20$. Note the log-scale on the y-axis in (a); the dashed line is a guide to the eye to emphasize the exponential dependence. (d)-(f) Dependence on system width for (d), (e) $L=40$ and (f) $L=20$. Parameters: $E_F=0.1$, $\lambda=t_3=0.1$. While the observed feature is nearly independent of the system width, the conductance minimum decreases exponentially with system length. The maximum conductance is nearly independent of the system length and the deviation from the quantized value presumably results from backscattering at the interface between leads and disordered region.}
\label{fig:sizedependence}
\end{figure}

\subsubsection{Dependence on system size}

In general, increasing the width of our strip has two opposing
effects. One, the overlap of edge states on opposite sides of the
strip is reduced and thereby the conductance should increase. Two, a
wider strip hosts more bulk channels and leakage from the edge into
the bulk is increased which reduces the conductance. Increasing the
length of our strip increases the probability of electrons inside the
wire being backscattered or leaking into the bulk since they travel a
longer distance inside the disordered region. This should decrease the
conductance while the net effect of changing the system width is not
apparent.

The dependencies of the characteristic values of the conductance, the
minimum $g_{\textnormal{min}}$, the maximum $g_{\textnormal{max}}$ at
higher disorder, and the disorder strength at the minimum $W_0$
(cf. Fig~\ref{schematicdip}), on the system size are given in
Fig.~\ref{fig:sizedependence}. We find that these quantities have rather different dependence on the system dimensions. $g_{\textnormal{min}}$ shows an exponential decay with increasing system length and a small increase with increasing
width. The disorder strength $W_0$ at the minimum seems to be independent of the
system width but shifts to smaller disorder values with increasing
system length. $g_{\textnormal{min}}$, $g_{\textnormal{max}}$, and $W_0$ are (nearly)
independent of the strip width which shows that either the two opposing
effects of changing the width are not very significant or that they (nearly) cancel each other for the
studied parameter region.

Rather importantly, it appears that the most conducting channel
recovers at high disorder and becomes effectively ballistic. The
conductance maximum $g_{\textnormal{max}}$ seems to be
independent of the system length. Its deviation from the 
quantized value is presumably the result of backscattering at the interface between
leads and disordered wire rather than from effects inside the disordered region.

\subsubsection{Dependence on system parameters}

\begin{figure}[t]
	\centering
		\includegraphics[width=6cm]{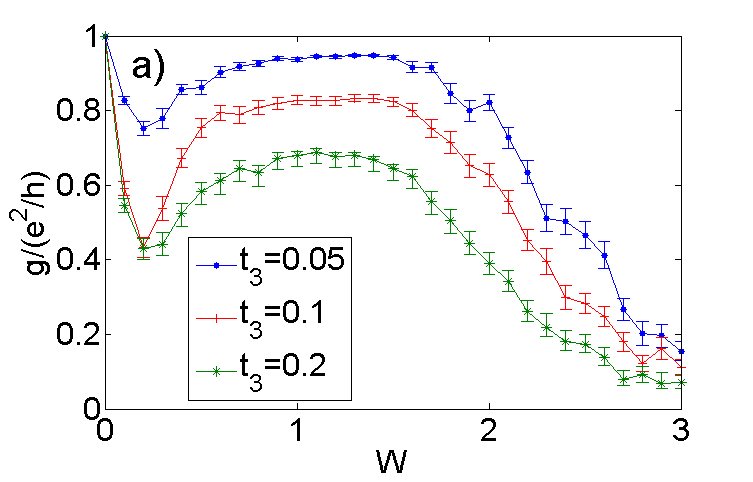}

\centering
		\includegraphics[width=6cm]{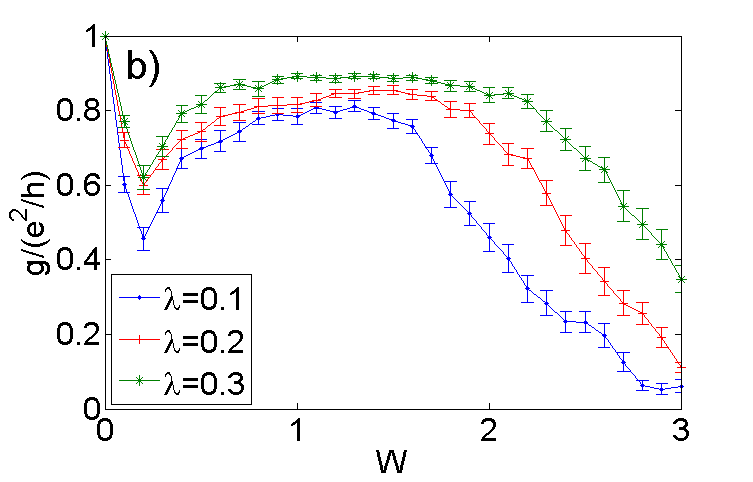}

	\centering
	
	\includegraphics[width=6cm]{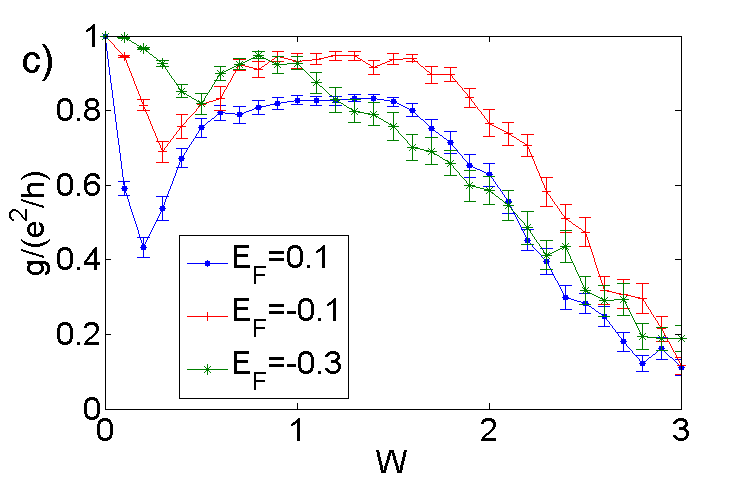}
	
	\centering
	\includegraphics[width=5cm]{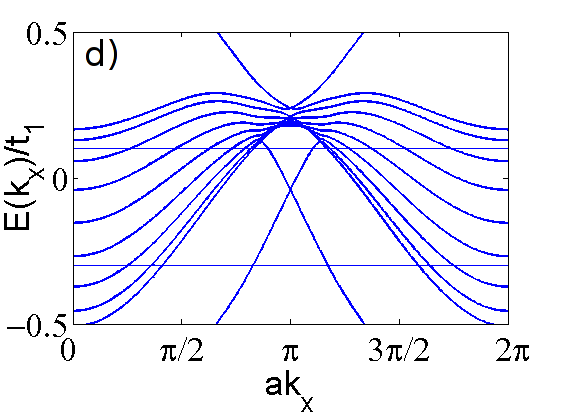}

\caption{Region I. Dependence of the edge channel conductance on system parameters. Dependence on (a) hybridization $t_3$ ($\lambda=0.1$, $M=20$) and (b) second-nearest-neighbor hopping $\lambda$ ($t_3=0.1$, $M=10$) for $E_F=0.1$ and $L=20$. (c) Dependence on Fermi energy with $\lambda=t_3=0.1$, $M=10$, and $L=20$; (d) indicates the energy range from $E=0.1$ to $E=-0.3$. Decreasing the hybridization $t_3$ and increasing the second-nearest-neighbor hopping $\lambda$ increases the conductance over the entire disorder range. Moving the Fermi energy away from the metallic bulk has the same effect until $E_F$ is too close the honeycomb valence band when the characteristic feature is starting to vanish.}
\label{paramdep}
\end{figure}

An increase in the coupling $t_3$ between the honeycomb subsystem and
the triangular lattice increases the probability of electrons leaking
from the edge into the localized metallic bulk states which should
reduce the conductance of the chiral edge states. Imaginary
second-nearest-neighbor hopping $\lambda$ is the essential ingredient
for a topological phase as it opens a gap and produces chiral edge
states. An increase in $\lambda$ widens the gap and thereby makes the
system more robust against disorder\cite{KaneMele1} which should
increase the conductance and broaden the maximum.

The dependence of the edge channel conductance on the hopping
parameters $\lambda$ and $t_3$ is illustrated in
Fig.~\ref{paramdep}(a) and (b). We find that the characteristic shape of the
conductance is left unaffected by changes in the hopping parameters of
over a wide range of values. As expected, an increase in the
hybridization $t_3$ decreases the conductance while an increase in the
second-nearest-neighbor hopping $\lambda$ increases it and leads to a broader maximum due
to the larger bulk gap of the honeycomb subsystem. One might think that the dependence of the conductance maximum on the hybridization $t_3$ contradicts our earlier conclusion that the deviation of the conductance maximum from the quantized value results from backscattering effects at the interface between conductor and leads. This is, however, not the case since an increasing coupling between the triangular and honeycomb lattices increases backscattering everywhere, including at the interface between conductor and leads, leading to the observed behavior.

When varying the Fermi energy within region I to values away from the
metallic band (see Fig.~\ref{paramdep}(c)), first the conductance of the most conducting channel increases over the entire
disorder range because of a decreasing number of metallic states but
the characteristic shape is unaffected. However, for Fermi energies
too close to the bulk of the honeycomb lattice, the conductance starts
to lose its shape. Even small disorder can in this case destroy to
topological bands through partial closing of the bulk gap of the
honeycomb subsystem.

\subsection{Region II}

\begin{figure}[t]
\begin{minipage}{4.2cm}
	\centering
	\includegraphics[width=4.5cm]{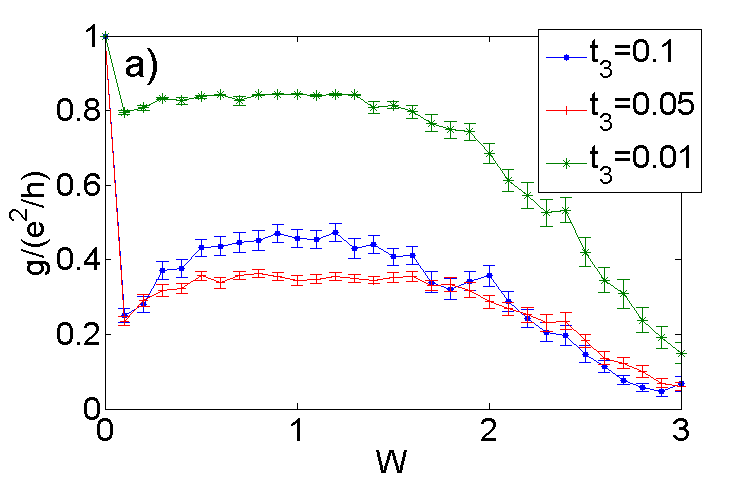}
\end{minipage}
\hfill
\begin{minipage}{4.2cm}
	\centering
	\includegraphics[width=4.5cm]{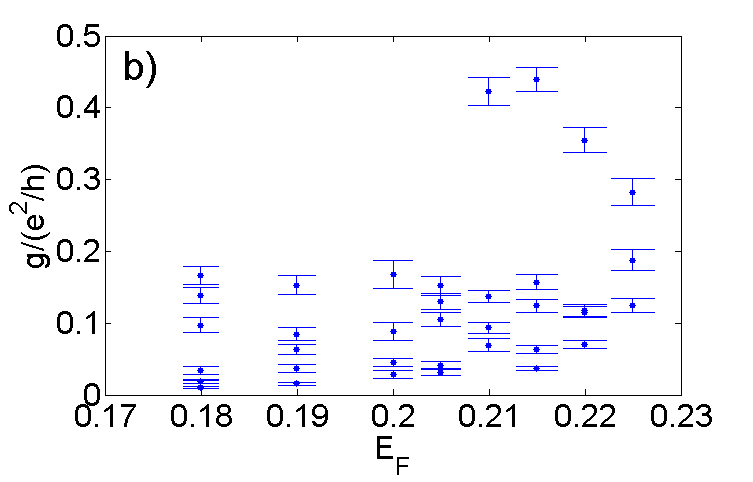}
\end{minipage}
\begin{minipage}{4.2cm}
	\centering
	\includegraphics[width=4.5cm]{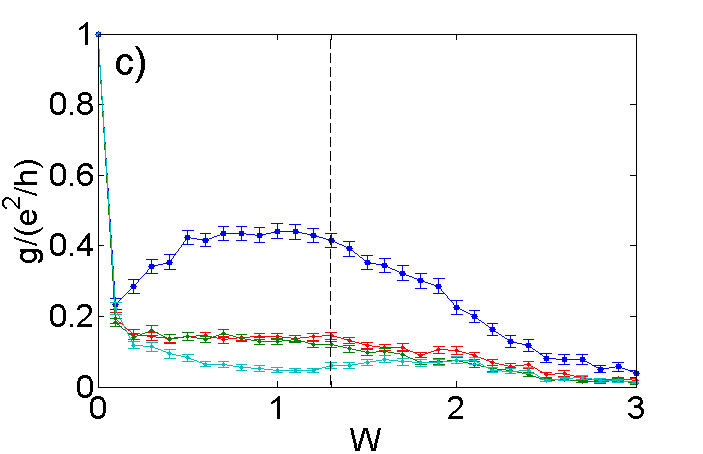}
\end{minipage}
\hfill
\begin{minipage}{4.2cm}
	\centering
	\includegraphics[width=4.3cm]{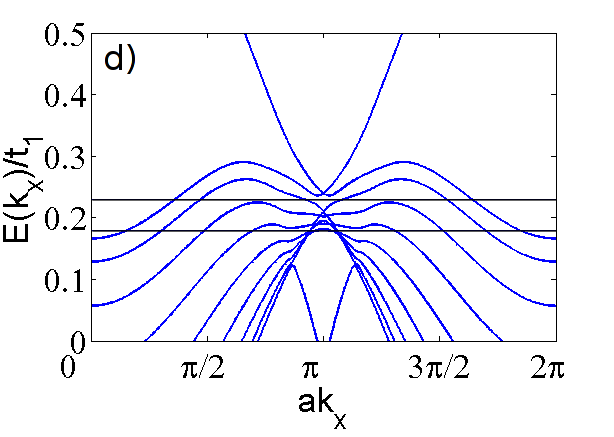}
\end{minipage}
	\caption{Region II. Conductance of most conducting channel for a Fermi energy deep within the metallic band. (a) Dependence on disorder strength for different hopping strengths $t_3$. $E_F=0.2$, $\lambda=0.1$, $M=20$, and $L=20$. (b) Conductance of each channel at different Fermi energies. For Fermi energies between $E=0.21$ and $E=0.22$ a single highly conducting channel exists possibly connected to surface resonances. Parameters are $W=1.3$, $\lambda=t_3=0.1$, $M=10$ and $L=20$. (c) Conductance as a function of disorder at $E_F=0.21$ for $\lambda=t_3=0.1$, $M=10$ and $L=20$. A single channel stands out, highly conducting at finite disorder. Dashed line indicates the disorder strength taken in (b). (c) shows the energy range from $E=0.18$ to $E=0.23$.}
\label{region2}
\end{figure}

In this region, the Fermi energy lies deep within the (trivial) metallic
band. Without hybridization between the honeycomb and the triangular
lattice, the edge states would traverse the bulk gap unhindered through
the metallic band. For finite hybridization, however, the edge states
are expelled out  of the metallic band region to higher and lower energies 
(cf. Fig.~\ref{spectrum1}). We would
therefore expect the conductance in this regime to show bulk state behavior, i.e.,
rapid decay with increasing disorder. Decreasing the hybridization
should recover the characteristic edge state behavior. We find,
however, that even for strong hybridization $t_3=0.1$, for a certain range of 
parameters there is a conducting channel which shows the
characteristic behavior of the edge channel (Fig.~\ref{region2}). This
highly conducting channel exists only for a range of Fermi energies within
the metallic band.  

The explanation of the highly conducting channel in this regime is
consistent with the notion of leftover spectral resonances of the edge
states, once they are absorbed by the bulk. When the surface bands are pushed away from the metallic band they
leave behind a finite lifetime surface resonance in place of the
original surface states (these are referred to as 'ghosts' in Ref.~\onlinecite{BergmanRefael}). 

\subsection{Region III}

\begin{figure}[t]
	\begin{center}
	\includegraphics[width=9cm]{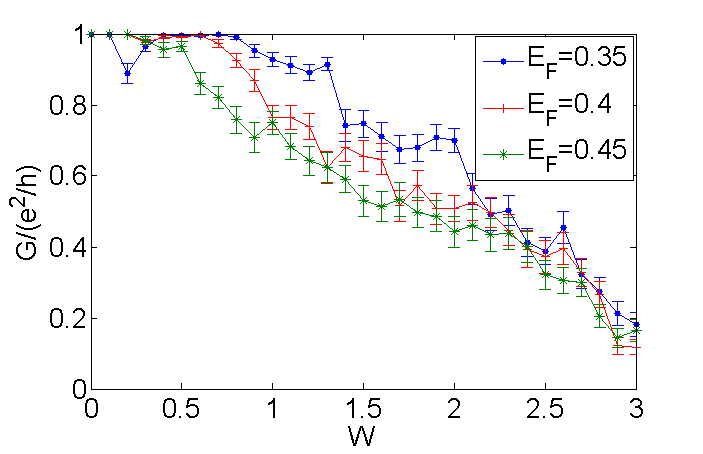}
\end{center}
\vspace{-3.5cm}
\hspace{-2.75cm}
	\includegraphics[width=3.2cm]{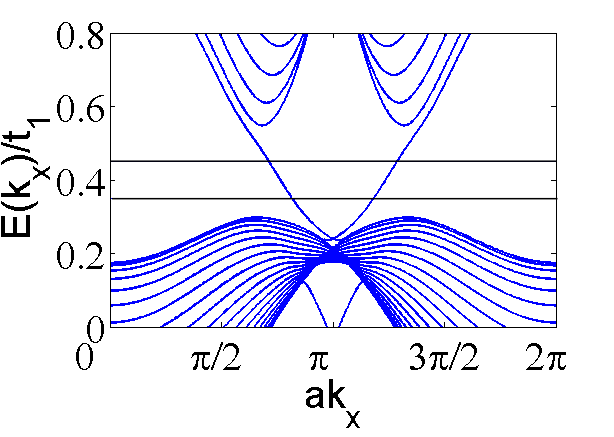}
\vspace{0.9cm}
	\caption{Region III. Conductance of the edge state for different Fermi energies. Parameters are $\lambda=t_3=0.1$ and $M=L=20$. Inlet indicates the energy range form $E=0.35$ to $E=0.45$. At Fermi energies close to the metallic bulk ($E_F=0.35$), disorder induced states above the bulk band still lead to a dip in the conductance. The closing of the honeycomb bulk gap due to disorder quickly decreases the conductance in the entire energy range.  }
\label{region3}
\end{figure}

In this region the Fermi energy lies above the metallic band and below
the conduction band of the honeycomb subsystem. The conductance is
only determined by the edge states and we would expect a quantized
value up to a certain critical disorder strength as in the case
without a metallic band. However, for Fermi energies very close to the
metallic band there is a residual effect from the metallic bulk states
and the conductance shows a small dip in the otherwise quantized
conductance (Fig.~\ref{region3}). 

Disorder closes the bulk gap of the honeycomb subsystem by creating
new bulk states above the valence and below the conduction band, and
the conductance decreases rapidly with disorder for larger Fermi
energies. In addition, disorder also creates new metallic states above
the metallic band. These new states then coexist with edge states as
in region I which explains the observed small dip at small disorder.

\subsection{Rashba coupling (Region I)}

\begin{figure}[t]
\begin{minipage}{4.2cm}
	\centering
	\includegraphics[width=4.4cm]{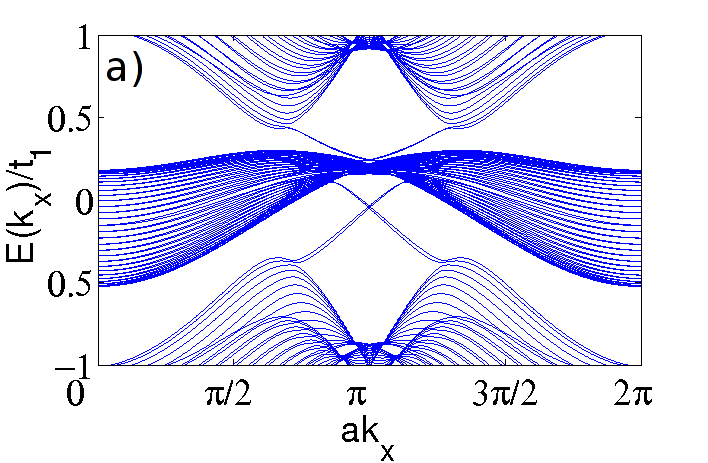}
\end{minipage}
\hfill
\begin{minipage}{4.2cm}
	\centering
	\includegraphics[width=4.4cm]{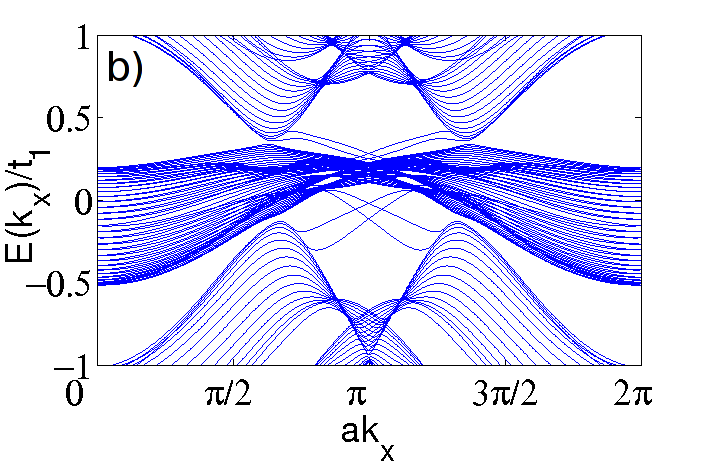}
\end{minipage}
	\caption{Band structure with Rashba coupling (a) $\lambda_R=0.1$ and (b) $\lambda_R=0.3$. Parameters are $\lambda=t_3=0.1$, and $M=30$. Rashba coupling reduces the band gap of the honeycomb subsystem.}
\label{rashbaspec}
\end{figure}

\begin{figure}[t]
	\centering
	\includegraphics[width=6.5cm]{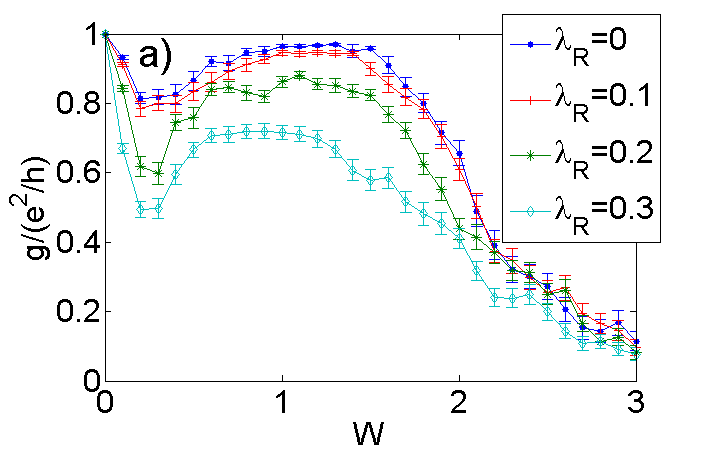}

	\centering
	\includegraphics[width=6.5cm]{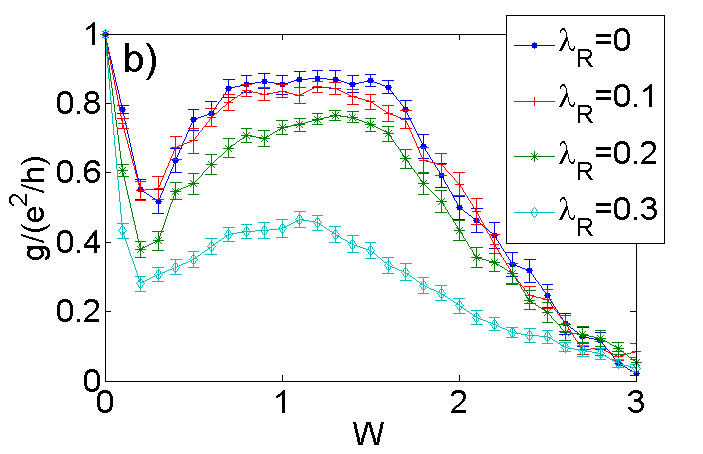}
	
	\caption{Region I with Rashba coupling. Conductance of one edge channel for (a) $t_3=0.05$ and (b) $t_3=0.1$ for different Rashba coupling strengths $\lambda_R$. Parameters are $E_F=0.05$, $\lambda=0.1$, $M=10$, and $L=20$. We plotted only a single spin orientation for readability. Rashba coupling overall reduces the conductance. The effect is enhanced for larger hybridization between the honeycomb and metallic subsystems.}
\label{rashbadep}
\end{figure}

We also considered a spinful version of \eqref{Hamiltonian} with Rashba type spin-orbit coupling included between nearest-neighbor sites of the honeycomb subsystem \eqref{Hrashba}. We considered the effects of spin-independent disorder only. As shown in Fig.~\ref{rashbaspec}, aside from the
doubling of the bands, small Rashba coupling does not change the band
structure dramatically. For large Rashba coupling the edge bands are clearly
separated in momentum and the structure of the metallic band changes
significantly. The bulk gap of the honeycomb subsystem is reduced such
that region III vanishes. However, we find that Rashba coupling does
not affect the characteristic shape of the edge channel conductance as
seen in Fig.~\ref{rashbadep}. Increasing Rashba coupling decreases the
conductance over the entire disorder range but leaves the distinct
'dip and recovery' behavior of the conductance intact. As for the model without Rashba
coupling, decreasing the hybridization $t_3$ between honeycomb and
triangular lattice increases the conductance.


\section{Interpretation and toy model\label{analytics}}

The nature of the conductance dip found in the previous section can be
qualitatively understood using a simple toy model. Our toy model
consists of a single chiral mode that is mixed, due to disorder, with
multiple bulk modes. This model neglects inter-edge scattering. We
further approximate the most conducting channel from one side to the
other as the sum of two incoherent contributions: edge-to-edge 
$g_{e-e} = \frac{e^2}{h} |T_{e,e}|^2$ (through just the edge channel) and bulk-to-edge 
$g_{e-b} = \frac{e^2}{h} \sum_{m \neq e} |T_{e,m}|^2$ transport. 

The transmission coefficient between mode $m$ in the left lead to mode $n$ in the right lead, can be computed from the retarded Green's function connecting these two modes \cite{Fisher1981}
\beq
T_{nm} = i \hbar \sqrt{v_n v_m} {\mathcal G}_{nm}
\; ,
\label{transg}
\eeq
where $v_n$ is the electron velocity in mode $n$ at the Fermi energy. For the sake of brevity we will take $\hbar \equiv 1$ from this point forward.

Indeed, since this system has a chiral
mode, strictly speaking, it will always have a total conductance bigger than
the single channel, $e^2/h$. Nevertheless, we will only be interested in
the conductance involving the bare chiral edge states. So if the
chiral mode manages to wander deep into the bulk due to the disorder,
we will ignore its conductance. This is consistent with the
numerics, since under these circumstances one would no longer be
able to separate the system into two independent parts, each
involving a single chiral mode. 

In the following we study the toy model, explicitly calculate the
conductance for different system lengths and widths, and use the
results to explain the origin of the conductance dip.

\subsection{Toy model and Green's functions}

Consider a system with a single chiral edge mode and multiple bulk
modes. We assume that the chiral edge mode can only hybridize with the
bulk when impurities are introduced. The hybridization strength will
serve as a proxy for disorder strength. 

We describe our system with the action
\beq
\begin{split}
S = & \int \frac{d\omega}{2\pi} dx dx' \Big[
\chi^{\dagger}(x) G^{-1}(x-x') \chi(x')
\\ & +
\psi^{\dagger}_{n}(x) F^{-1}_{n m}(x-x') \psi^{\phantom\dagger}_m(x')
\\ & -
\left[ \chi^{\dagger}(x) V^{\phantom\dagger}_n(x) \psi^{\phantom\dagger}_n(x) + h.c. \right]
\Big]
\; ,
\end{split}
\eeq
where the field $\chi(x)$ represents an electron at position $x$ in the bare edge mode, and the fields $\psi_n(x)$ represent an electron at $x$ in bulk mode $n$. In addition, $G$ and $F$ denote the bare edge and bulk state propagators, respectively, and $V$ denotes the random potential scattering between the edge state and bulk channels. We will adopt the uniform distribution of random potential strength. The bare propagators are translationally invariant in the direction through the length of the strip (hence the dependence only on the relative coordinate $x-x'$), while the impurity scattering $V(x)$ is completely local. All terms depend on frequency $\omega$, which for the sake of brevity has been kept implicit. The Einstein summation convention has been used for the bulk channel indices $n,m$. In momentum space the form of the propagators is
\beq
\begin{split} &
G^{-1}(k,\omega) = \omega + i \eta - v_e k
\\ & 
F^{-1}_{n,m}(k,\omega) = \delta_{n,m} \left[ \omega + i \eta - \frac{k^2}{2 {\mathcal M}_n} + \mu_n \right]
\; ,
\end{split}
\eeq
where $\eta>0$ is infinitesimal and positive, and ${\mathcal M}_n$ and $\mu_n$ are the effective mass and the chemical potential, respectively, in bulk mode $n$. Fourier transforming from momentum space to real space we find
\beq\label{ansatz0}
\begin{split}
G(x,E_F) & = \int_{-\infty}^{+\infty} G(k,E_F) e^{i k x} = 
-\frac{i}{v_e} e^{i k_e x} \theta(x)
\\ 
F_{n,m}(x,E_F) & = \int_{-\infty}^{+\infty} F_{n,m}(k,E_F) e^{i k x}\\
& = 
\frac{-i}{k_{b,n}/{\mathcal M}_n} e^{i k_{b,n} |x|} \delta_{n,m}
\; ,
\end{split}
\eeq
where $k_{b,n} = \sqrt{2 {\mathcal M}_n (E_F+\mu_n)}$ is the Fermi momentum in the bulk mode $n$, and $k_e = \frac{E_F}{v_e}$. The chiral mode propagator $G$ reflects a single propagation direction. $G(x)$ is zero when $x<0$, and this reverses when we take $\eta<0$ for the advanced propagator. For the bulk modes, we assume $F$ is
diagonal, neglecting the mixing of different bulk conduction channels. 

The quantity $V(x)$ is not the bare random potential, but rather matrix elements between bulk modes and the edge state, due to disorder scattering, and we therefore assume they are uncorrelated random complex variables. The random phase captures the random location of the focal points of impurity induced edge-bulk scattering, and makes sure our results are not affected by momentum conservation. In this manner we take into account the diffusive nature of the bulk, while still using the translationally invariant form of the bulk propagator $F$.

\begin{figure}[h]
\begin{center}
\includegraphics[width=85mm]{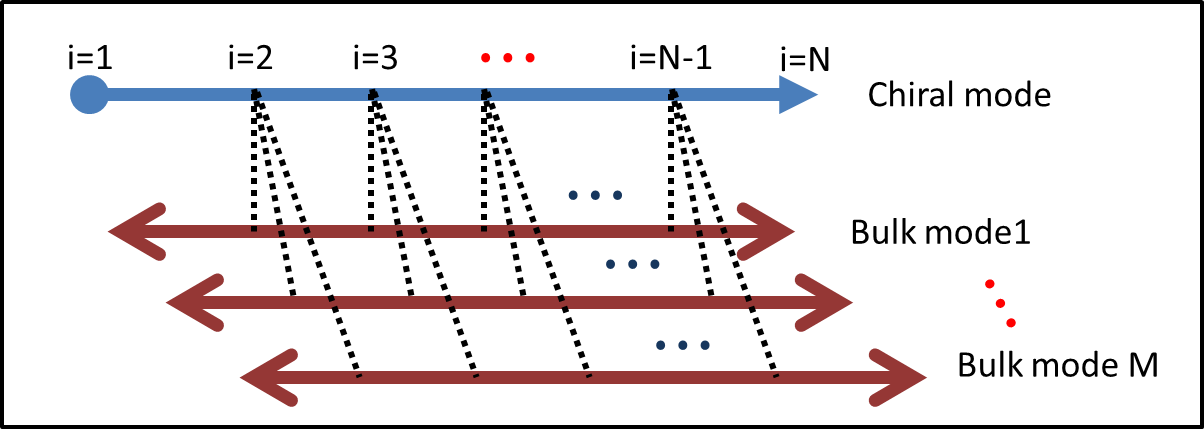}
\caption{Illustration of the toy model. Bulk modes are independent
  of each other since we assume that the bulk channels describe the
  eigenmodes of the diffusive bulk. The edge mode couples to the bulk
  through $(N-2)$ scattering sites, giving in total $(N-2)M$
  couplings $V_n(i)$. The coupling strength and phase are uniformly distributed.}
\label{tmfig1}
\end{center}
\end{figure}

We discretize the $x$ coordinate into $i=1 \ldots N$, as illustrated in Fig.~\ref{tmfig1}, and set $V_n(i=1) = V_n(i=N) = 0$ to represent clean leads so that we can still separate in our calculations the edge conduction channel from all the others. For $i \neq 1,N$ we take $V_n(j) = t_{j,n}e^{\phi_{j,n}}$. Now $G$ and $F$ are both square matrices with sizes $N \times N$ and $N M \times N M$, respectively, where $M$ is the number of channels (and the width of the lattice model of our strip). In contrast, $V_n$ is a non-square matrix of dimensions $N \times N M$. Discretizing $x$ requires one subtle change to the chiral propagator $G$ in \eqref{ansatz0}
\beq\label{ansatz}
\begin{split} &
G(x_i - x_j,E_F) =
-\frac{i}{v_e} \left[
e^{i k_e (x_i - x_j)} \theta(x_i-x_j)
+ \frac{1}{2} \delta_{i j}
\right]
\\ & 
F_{n,m}(x_i-x_j,E_F) = 
\frac{-i}{k_{b,n}/m_n} e^{i k_{b,} |x_i-x_j|} \delta_{n,m}
\; ,
\end{split}
\eeq
where the step function $\theta (x)$ at $x=0$ is defined to be zero.
To find the transmission coefficients we seek, we will need to calculate the Green's function for electrons starting out at the left lead $i=1$ and coming out at the right lead $i=N$, at the Fermi energy
\beq
\begin{split} &
{\mathcal G}_{e,e}(x = N, x' = 1) = \langle \chi_{i=N} \chi_{i=1} \rangle
\\ & 
{\mathcal G}_{e,m}(x = N, x' = 1) = \langle \chi_{i=N} \psi_{m,i=1}  \rangle
\; .
\end{split}
\eeq
where we specifed two contributions: electron from edge to edge mode, ${\mathcal G}_{e,e}$, and electron from bulk mode m to edge mode, ${\mathcal G}_{e,m}$. With the position index suppressed for the sake of brevity, our action is of the form
\beq
S = \int d\omega  \left[  \begin{pmatrix} \chi^\dagger & \psi^\dagger \end{pmatrix}  \begin{pmatrix} G^{-1} & -V \\ -V^\dagger & F^{-1} \end{pmatrix} \begin{pmatrix} \chi \\  \psi \end{pmatrix} \right]\label{Htop}
\; .
\eeq
Next we want to invert the matrix, to find the components of the full Green's function 
${\mathcal G} = \begin{pmatrix} G^{-1} & -V \\ -V^\dagger & F^{-1} \end{pmatrix}^{-1}$.
We find
\beq\label{g1}
\begin{split} &
{\mathcal G}_{e,e} = 
\frac{1}{G^{-1} - V F V^\dagger} 
\\ &
{\mathcal G}_{e,m} = {\mathcal G}_{e,e} V_n F_{n m} = {\mathcal G}_{e,e} V F
\\ & 
{\mathcal G}_{m,e} = {\mathcal G}_{m,n} V_n^* G
\\ & 
{\mathcal G}_{n,m} = \left( \frac{1}{F^{-1} - V^\dagger G V } \right)_{n,m}
\; ,
\end{split}
\eeq
where we have used the Einstein summation convention.
The non-square matrices ${\mathcal G}_{e,m}$ and ${\mathcal G}_{m,e}$ are the Green's function elements that capture the bulk-edge and edge-bulk mixing, respectively; ${\mathcal G}_{e,e}$ and ${\mathcal G}_{n,m}$ are the renormalized Green's functions for the edge and bulk modes, respectively. 

As we will see in what follows, the simple toy model already affords us a
qualitative understanding of the conductance dip, and its dependence on the system dimensions.

\subsection{Conductance dip of different system size}

\subsubsection{Single contact with single bulk case}

It is instructive to first consider the case of a single bulk mode $m=b$ and
a single impurity. In this case we take $N=3$ and $M=1$, and for simplicity we take $k_e = k_b = k_F$, and $V = t e^{i \phi}$. From Eqs. (\ref{ansatz},\ref{g1}), we can show that
\beq
\begin{split} &
{\mathcal G}_{b,b}(N,1) = \frac{e^{i k_F L}}{v_b}\frac{1}{1+t'^2/2}
\\ &
{\mathcal G}_{e,e}(N,1) =  \frac{e^{i k_F L}}{v_e}\frac{1-t'^2/2}{1+t'^2/2}
\\ &
{\mathcal G}_{b,e}(N,1) = \frac{e^{i k_F L}}{\sqrt{v_ev_b}}\frac{t'}{1+t'^2/2}
\\ &
{\mathcal G}_{e,b}(1,N) = \frac{e^{i k_F L}}{\sqrt{v_ev_b}}\frac{t'}{1+t'^2/2}
\; ,
\end{split}
\eeq
where $L=(N-1)a = 2 a$, $a$ is the lattice spacing, and $t'=t/\sqrt{v_e v_b}$ is a dimensionless coupling strength. From these we find the conductances
\beq
\begin{split} &
g_{e-e} = \frac{e^2}{h}
\left[
\frac{1-t'^2/2}{1+t'^2/2}
\right]^2
\\ & 
g_{e-b} = \frac{e^2}{h}
\left[ 
\frac{t'}{1+t'^2/2} 
\right]^2
\; .
\end{split}
\eeq
We can now easily see that $g_{e-e}=\frac{e^2}{h}$ for both the clean limit $t'=0$ and the infinite disorder limit $t' \rightarrow \infty$. In between there is a minimum of $g_{e-e}=0$ at $t'=\sqrt{2}$.
In contrast, $g_{e-b}$ starts at zero for $t'=0$, and rises to a maximum value of $g_{e-b} = \frac{1}{2} \frac{e^2}{h}$ at $t'=\sqrt{2}$, before decaying back down to zero. Note that the Fermi velocities only rescale the disorder strength, and did not otherwise influence the conductance behavior.

\begin{figure}[h]
\begin{center}
\includegraphics[width=85mm]{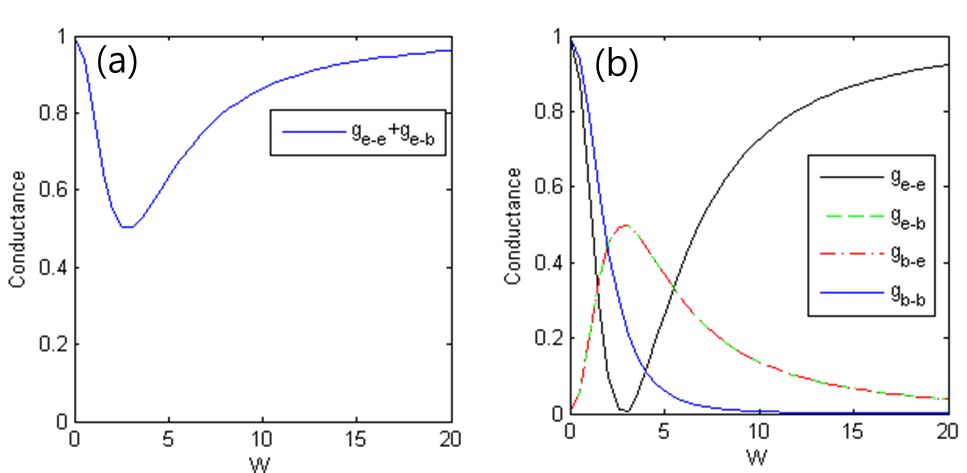}
\caption{Conductance contribution via different channels as a function
  of coupling strength (x-axis) for the system of a single chiral and
  bulk mode with a single coupling: from edge to edge $g_{e-e}$, edge to
  bulk $g_{b-e}$, bulk to edge $g_{e-b}$, and bulk to bulk $g_{b-b}$. The edge to edge conductance has a minimum at finite disorder before recovering to the quantized value. The conductance is in units of $\frac{e^2}{h}$.}
\label{tmfig2}
\end{center}
\end{figure}

Finally using Eq. (\ref{transg}) we find all the different contributions to the conductance (edge/bulk-edge/bulk), as plotted in Fig. \ref{tmfig2}.
Fig. \ref{tmfig2}(a) shows the total conductance of the strip,
consisting of adding up the edge-to-edge and bulk-to-edge
channels. Fig. \ref{tmfig2}(b) shows the possible individual
conductances. Two points are noteworthy: (i) In the strong disorder
limit, the conductance of edge-to-edge channel converges to
unity. (ii) At the conductance dip minimum, the biggest contribution
to the conductance comes from the  bulk-edge channel. In other words,
at strong disorder, the edge-edge channel conductance recovers to unity even without accounting for the localization of the
bulk, and the size dependence of the conductance near the dip will arise from the
bulk-edge channel contribution, and will be rather independent of the
edge-edge channel behavior.

\subsubsection{Multiple bulk modes and impurities}

Consider now a general system of length $N$ (i.e., having $N-2$
scattering sites), and width (bulk mode number) $M$. The Green's function components we need to calculate to find the transmission coefficient ${\mathcal G}(N,1)$ can be found in closed form. In particular, the edge-edge Green's function is found to be
\beq
{\mathcal G}_{e,e}(N,1) = -i e^{i k_e (N-1)} \frac{det( G^T - \vv F \vv^{\dagger})}{det(G + \vv F \vv^{\dagger})} 
\; ,
\label{g15}
\eeq
where $\vv_n(x_j) = (-1)^j e^{i k_e x_j} V_n(x_j)/2 $, such that $ (\vv F \vv^{\dagger})_{i,j} = \sum_{n,m} \vv_n(i) F_{n,m}(i,j) \vv_m(j)^*$. The derivation of this result is somewhat involved, and so we defer its details to a future publication.\cite{KimJunckKlichRefael} More generally, for $j \neq 1$, we can show that
\beq
{\mathcal G}_{e,e}(N,j) = i\frac{(-1)^j}{2} e^{i k_e(N-j)}  
\frac{det(G^T - \vv F \vv^\dagger)_{j^{th}}}{det(G + \vv F \vv^\dagger)}  \label{g17}
\eeq
where $(A)_{j^{th}}$ indicates the $j^{th}$ row of matrix is
replaced by a vector $(a)_j=-i(1-\delta _{1,j}-\delta _{N,j}) = (0,-i,-i,\ldots,-i,0)$.
Concomitantly, the Green's function bulk-edge mixing component is
\beq
{\mathcal G}_{e,m}(N,1) = \sum_n \sum_{j=2}^{N-1} {\mathcal G}_{e,e}(N,j) V_n(j) F_{n,m}(j,1)
\eeq
which is the sum of  the $(N-2)$ Green's functions from clean bulk to
renormalized chiral mode. 

\subsubsection{Length dependence}

\begin{figure}[t]
\begin{center}
\includegraphics[width=85mm]{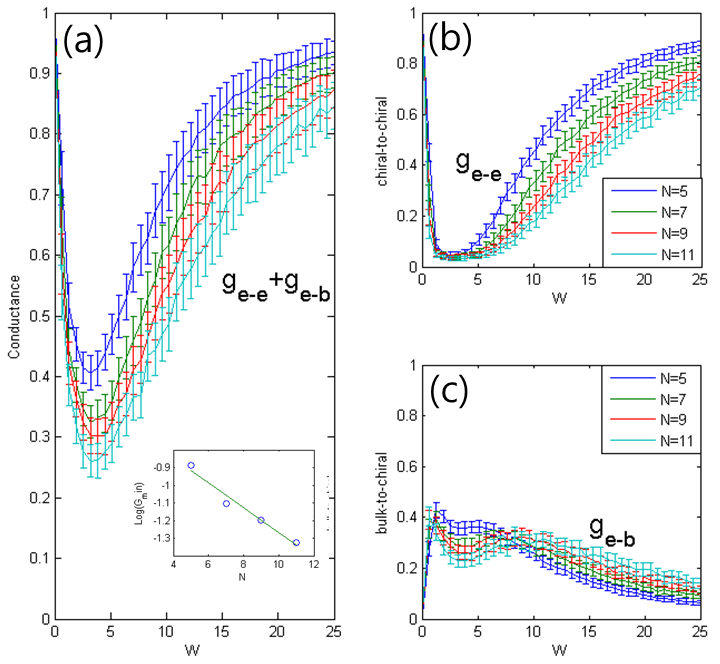}
\caption{(a) Sum of the conductances of chiral-to-chiral (b) and bulk-to-chiral (c) channels for M=4 and N=5 (blue), 7 (green), 9 (red) and 11 (cyan) for $k_e=\frac{1.4\pi}{a}$ and $k_b=\frac{1.2 \pi}{a}$. The inset in (a) shows the exponential decay of the minimum conductance with system length in agreement with the numerical calculation in Fig.~\ref{fig:sizedependence}(a). The conductance is in units of $\frac{e^2}{h}$.}\label{tmfig3}
\end{center}
\end{figure}

The conductance minimum $g_{\textrm{min}}$ is expected to drop exponentially with the system's length, as shown in
Fig. \ref{fig:sizedependence}(a). The toy model's results are shown in Fig.\ref{tmfig3}.  Within our toy model, the length dependence of $g_{\textrm{min}}$
is attributed to the mixed edge-bulk contribution to the conductance $g_{e-b}$. Indeed,
we approximate the edge-related conduction as an incoherent sum of two
contributions: conductance of electrons entering and leaving the strip
in an edge state, $g_{e-e}$, and entering as bulk states, but leaving the
strip as edge electrons, $g_{e-b}$. Even with multiple scatterers, $g_{e-e}$
behaves as Eq. (\ref{g15}) indicates: it drops down from unity (in units of $\frac{e^2}{h}$) to zero at moderate disorder, then rises back to unity at strong
disorder. In Fig. \ref{tmfig3}(b) we see that the minimum $g_{e-e}$ is independent of the system length. Therefore, the only possible source of the $g_{\textrm{min}}$ dependence on the system length must come from $g_{e-b}$, which starts at zero for the clean limit, rises for moderate disorder, then peaks and decays as we go to strong disorder
(see Fig. \ref{tmfig3}(c)). The inset in Fig. \ref{tmfig3}(a) shows the
conductance $g_{e-b}$ depends exponentially on the number of scattering
points, and hence on the length of the system.

\subsubsection{Width dependence}

We now explore the width dependence of the conductance through the edge states. To this end, we can use a simple approximation $\vv F \vv^{\dagger} \approx \left<\vv F \vv^{\dagger}\right>$. This stems from the fact that $\left<\left( \vv F \vv^{\dagger}\right)^n\right>/\left(\left< \vv F \vv^{\dagger}\right>\right)^n \rightarrow 1$ when the number of bulk modes goes to infinity $M \rightarrow \infty$. This holds as long as the distribution of the random matrix elements $V_{j,n}$ has a finite variance. Our approximation is therefore very good when there is a large number of bulk modes $M \gg 1$, and we can then write
\beq 
\left< {\mathcal G}_{e,e}(N,1) \right> \simeq  -i e^{i k_e (N-1)} \frac{det( G^T - \left< \vv F \vv^{\dagger}\right> )}{det(G + \left<\vv F \vv^{\dagger}\right> )} 
\; .
\eeq
A similar expression holds for ${\mathcal G}_{e,e}(N,j\neq 1)$ as well.

The width dependence of the conductance can now be deduced from
\beq
\begin{split} &
\left( \left< \vv F \vv^{\dagger}\right> \right)_{ij} 
\\ & = 
(-1)^{i+j} e^{i k_e (x_i - x_j)} \sum_{n m} F_{n,m}(i,j) \left< V_n(i) V_m(j)^* \right>
\\ & = 
(-1)^{i+j} e^{i k_e (x_i - x_j)} \sum_{n m} F_{n,m}(i,j) \delta_{i j} \delta_{n m} W^2
\\ &
= 
\delta_{i j} W^2 \sum_{n} F_{n,n}(i,i)
\sim M \delta_{i j} W^2
\; ,
\end{split}
\eeq
where the variance of the disorder is the disorder strength $W$ squared.
From this we see that the dependence on $M$ can be absorbed into $W$,
\beq
W_{\textrm{eff}} = \sqrt{M} W
\; .
\eeq
As a consequence we expect that increasing the width (increasing $M$) only shifts the minimum to different disorder strength $W_0$, and does not change $g_{\textrm{min}}$.

\begin{figure}[t]
\begin{center}
\includegraphics[width=85mm]{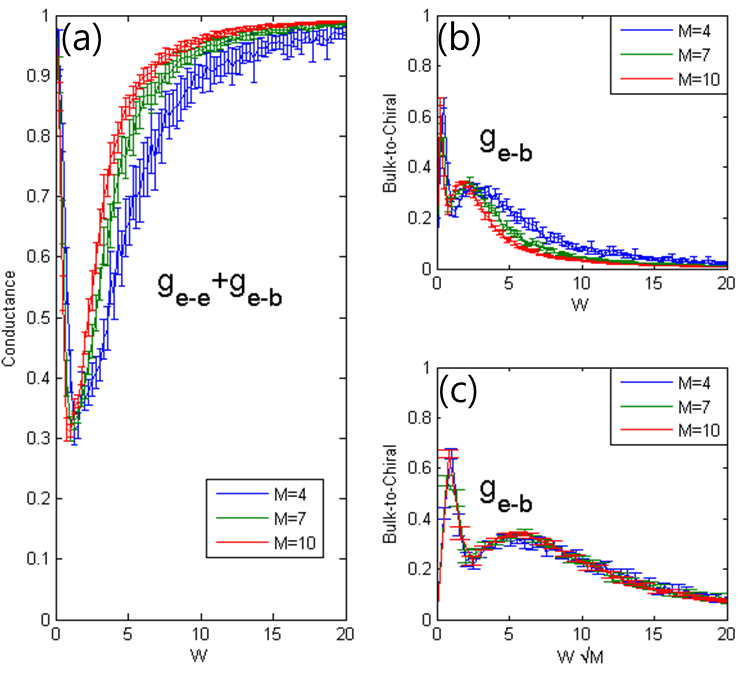}
\caption{(a) Sum of conductance of chiral-to-chiral (b) and bulk-to-chiral (c) channels for N=5 and M=4 (blue), 7 (green), 10 (red) for $k_e=\frac{1.4\pi}{a}$ and $k_b=\frac{1.2\pi}{a}$. The conductance minimum is unaffected by changes is the system width in agreement with the results of the numerical calculation. c) Renormalizing the disorder strength by $\sqrt{M}$, $g_{e-b}$, which is the dominant contribution to the conductance in the region of the minimum, becomes independent of the system width. The conductance is in units of $\frac{e^2}{h}$.}
\label{tmfig4}
\end{center}
\end{figure}

This simple dependence on $M$ turns out to work quite well, when applied to the toy model. Fig. \ref{tmfig4}(a) shows the conductance through the chiral mode for different system widths, $M$. The minimum conductances in the plots are indeed all similar, consistent with the numerical results shown in Fig. \ref{fig:sizedependence}(d). The conductance is dominated by the bulk-to-chiral channel (Fig. \ref{tmfig4}(b)). Furthermore, scaling the disorder strength by $\sqrt{M}$, the three lines of Fig. \ref{tmfig4}(b) collapse onto a single line in Fig. \ref{tmfig4}(c). Only the width of the minimum, and not its depth, is affected by the channel number.

\subsection{Summary of toy model analysis}

We have shown in this section that coupling between edge and
bulk modes leads in general to the conductance dip behavior with
increasing disorder strength. The dependence of the minimum conductance at the dip $g_{\textrm{min}}$ on the system size is consistent with the previous
calculation: it exponentially decays with the system length and is roughly independent of the system width. Nevertheless, the toy model is limited. First of all, the
localization of the whole system at strong disorder is not captured by it because the toy model assumes a fixed number of chiral and
bulk modes at a certain Fermi level. Second, the toy model is directly
concerned with the tunneling between modes and not equipped to
capture reflection within a single mode. As a consequence, the non-quantized value of the conductance plateau at strong disorder cannot be captured. Third, the toy model is
not suitable to describe the region II calculation where edge and bulk modes are coupled already in the clean limit, and as such are not well separated contrary to what we assume in the toy model.

\section{Conclusions\label{conc}}

In this manuscript we explored the transport signatures in a topological conductor phase.  The particular realization of a 2d topological conductor we considered here, had an edge state originating in the high-energy bands, overlapping in energy (but not in momentum) with a trivial band at the Fermi energy (see the spectrum in Fig.~\ref{spectrum1}). This model should suffice to capture the generic qualitative features of the non-protected topological aspects of topological conductors.

To find topological signatures, we calculated the two-terminal conductance through a topological conductor strip in the presence of disorder in the form of a random on-site chemical potential. Our results clearly show that even when a bulk gap is absent, the edge states, which are no longer protected, maintain their individuality and dominate the transport in the strip. Inspecting the effects of disorder on all the conducting channels of the topological conductor, a single channel clearly stands out (see Fig.~\ref{schematicdip}). The most conducting channel in the strip has a conductance which initially decays rapidly as we increase disorder strength, but then reaches a minimum, and then recovers to a value consistent with ballistic transport through the disordered region. The dip and recovery features single out the topological edge channel, and show that even when the system is gapless, its capacity for nearly dissipationless transport is restored once disorder suppresses transport through the bulk.  

The origin of the 'dip and recovery' features (Fig.~\ref{schematicdip}) presumably has to do with localization effects in the bulk. Intuitively, as the bulk states become less transparent, the edge state, which cannot localize since it is chiral, reemerges to dominate transport. To gain a better qualitative understanding of the main features in our numerics, we developed a simple toy model, consisting of a single chiral mode that can randomly scatter to several regular channels. The model assumes that the bulk is diffusive, and does not invoke localization explicitly. Disorder strength is encoded in the random scattering strength between the chiral edge and the bulk. The diffusive nature of the bulk is taken into account by using random phases in the edge-bulk scattering elements. Despite its simplifications, this model demonstrates that generically the edge-edge scattering contribution to the conductance $g_{e-e}$ shows the 'dip and recovery' behavior (see Figs.~\ref{tmfig2},\ref{tmfig3}).

Thus it appears the full-fledged Anderson localization, which is necessary to eliminate conductivity in the bulk, is not necessary for the revival of the edge state conductance. Rather, the dip and recovery feature appears to be guaranteed once a separate chiral mode is present, assuming it is unable to scatter to its counter-propagating partner. In the numerics, we do have both edge channels, but because their wave functions essentially do not overlap, disorder can not produce direct scattering between them. Note, however, that it appears that the 'dip and recovery' feature may not occur when the edge state overlaps in energy, and therefore, can hybridize, with the bulk states of its progenitor topological band.

The unique transport features in topological conductor strips will have additional signatures. First, we note that the results we presented here, when applied to a spinful electron system, rely on time-reversal invariance, and require that no scattering occurs between the counter-propagating, opposite-spin modes on the same edge. Therefore, when moderate disorder is present, and the strip conductance is dominated by the descendant of the edge channel, the conductance of the strip would be strongly suppressed by a magnetic (Zeeman) field normal to the Rashba spin direction. Such a Zeeman field opens up a gap in the edge states, pushing them away from the Fermi energy, in which case the edge state no longer contributes to transport. The highly conducting mode we found would then no longer contribute to the overall conductance through the strip, which will now be purely diffusive, with no remnant of ballistic transport. Magnetic impurities at the edge would have a similar destructive effect. Second, probing the current density in the sample would disclose that in moderate disorder the conduction is dominated by regions near the edge. Probing these effects experimentally would rely on materials or quantum wells having a band structure similar to the one considered here: A metallic band coexisting at the same energy with the edge states of a different band. 

The 'dip and recovery' feature arising in transport due to disorder was shown here to appear in a non-interacting 2d system. It is interesting to ask how these features would change when considering Coulomb interactions. This could also be explored within the toy model we presented, by promoting the electronic modes to Luttinger liquids with interaction between the various modes. We could then also consider a more complete model containing the chiral modes on both sides of the strip, and follow the mixing between them explicitly. Similarly, one could ask about similar resurgence of edge physics in a 3d topological conductor. Instead of conductance, however, the appropriate
topological signatures in 3d should be the Kerr effect rotation angle, which encodes the unique response of topological edge states to an oscillating electro-magnetic field (see also Ref.~\onlinecite{Voijta}).  

The work presented in this manuscript adds to the notion that topological behavior has distinct and strong signatures even in the absence of protection due to a band gap. This observation, as articulated also in Refs.~\onlinecite{Bergman,BarkeshliQi,HastingsLoring2}, significantly extends the class of systems (and materials) in which topological effects could be found. 

We thank Felix von Oppen for helpful discussions. We would like to acknowledge financial support through the Helmholtz Virtual Institute ``New states of matter and their excitations'' (AJ and GR), from DARPA (KWK and GR), of the Sherman Fairchild Foundation (DLB), and from NSERC and FQRNT (TPB).

\appendix*

\section{Transfer-matrix method \label{appA}}

In this section, we will demonstrate a way to calculate the transmission coefficients $T^{LR}_{nm}$ of our system using the transfer-matrix method.\cite{Tsu}

\begin{figure}
 	\centering
 		\includegraphics[width=3.2in]{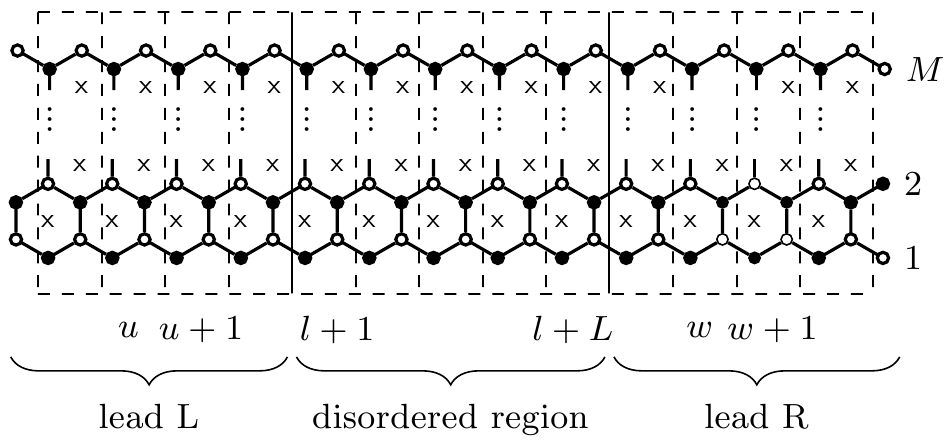}
\caption{Lattice system with a width of $M$ zig-zag lines. The disordered region of length $L$ is the conductor which is connected to a lead on either side. Dashed rectangles show the unit cells of the strip geometry.}
 	\label{stripfig}
 \end{figure}

Naively, all we have to do to calculate the transmission coefficients through our system is first find the basis of modes in the clean limit (the modes in the leads) $\psi_n(x)$, then starting with an electron in just one of the (clean-limit) modes in the left lead $\psi_n(x=a u)$ (cf. Fig.~\ref{stripfig}), propagate the wavefunction through the dirty sample, and calculate its weight in the various modes of the (clean) right lead $\psi(x=a w) = \sum_m T^{LR}_{nm} \psi_m(x=a w)$. These weights are precisely the transmission coefficients.

In practice, we run into some difficulties. Namely, the calculated $\psi(x=a w)$ contains weight in some unphysical exponentially growing modes $\sim e^{+|\kappa| x}$, not just in the propagating modes. As a consequence, we have to perform a more complicated numerical procedure.

\begin{figure}
 	\centering
 		\includegraphics[width=3.2in]{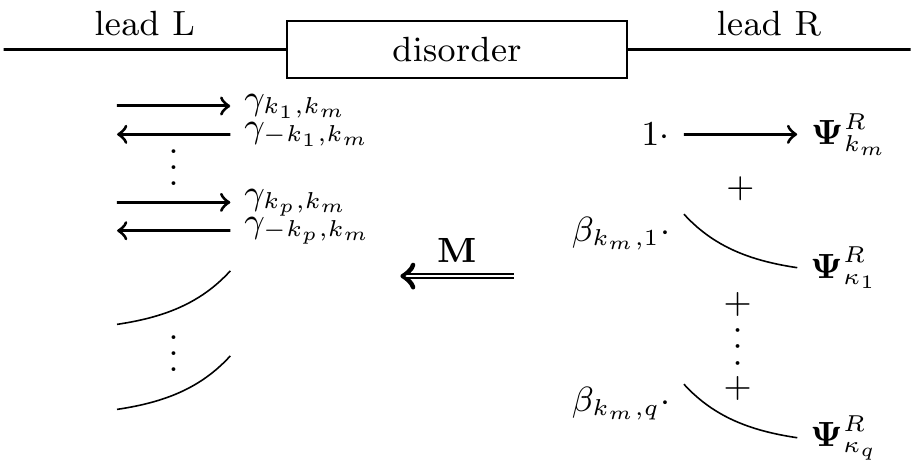}
\caption{Considering a superposition of physical modes in lead R, one can cancel the unphysical modes in lead L by destructive interference. The coefficients $\beta_{k_m,n}$ have to be chosen such that, after backwards propagation, the exponentially increasing modes in lead L have zero weight. The coefficients $\beta_{k_m,n}$ are given by Eq.~\eqref{eq:solve}. The coefficients $\gamma_{q^{}_n,q'_m}$ of the resulting superposition (of only physical modes) in lead L are given by Eq.~\eqref{eq:gamma}.}
 	\label{fig:schematic2}
 \end{figure}

The main steps of the calculation are the following: After determining the transfer matrix, we calculate the eigenvalues and -vectors of the transfer matrix which correspond to the possible modes and wavefunctions in the leads. For computational convenience we start the calculation in lead R. Propagating one of these wavefunctions backwards through the system using the transfer matrix gives us a wavefunction in lead L which is a superposition of all possible wavefunctions in the leads, i.e., a superposition of all eigenvectors of the transfer matrix including unphysical (exponentially increasing) modes. Next, in order to determine the transmission coefficients, we need to find a superposition of only physical modes in lead L which, after propagation through the system, leads to a superposition of one right-moving mode with weight one and decaying modes in lead R (cf. Fig.~\ref{fig:schematic2}). To find the necessary coefficients of the modes in lead L, we propagate a superposition of one right-moving mode with weight one and all decaying modes backwards through the system and choose the coefficients of the decaying modes such that the unphysical modes in lead L are canceled by destructive interference. Doing this for all right-moving modes gives us information about all transmission channels and we can determine the transmission coefficients $T^{LR}_{nm}$ for arbitrary modes $n$ and $m$ using probability current conservation.

The lattice model of the topological conductor is illustrated in Fig.~\ref{stripfig}. The system is a honeycomb strip with a width of $M$ zig-zag lines with an overlayed triangular lattice with sites that occupy the centers of the honeycomb hexagons. The unit cell is chosen as explained in Fig. \ref{stripfig} with dashed rectangles indicating the unit cells. 
A region with a length of $L$ unit cells constitutes the disordered topological conductor. It is connected to two perfect ballistic leads, one on each side. 


In order to determine the transfer matrix, the Hamiltonian given by Eq.~\eqref{Hamiltonian}, which only connects sites within neighboring unit cells (cf. Fig.~\ref{stripfig}), can be written as

\begin{equation}
{\cal H}=\sum_u{\mathbf{c}_{u}^{\dagger}(\mathbf{H}_{u,u-1}\mathbf{c}_{u-1} +\mathbf{H}_{u,u}\mathbf{c}_{u} +\mathbf{H}_{u,u+1}\mathbf{c}_{u+1}) },
\label{eq:hamgeneral}
\end{equation}
\noindent where \[\mathbf{c}_{u}^{\dagger}=\begin{pmatrix} (a_{u,1})^{\dagger}&&(b_{u,1})^{\dagger}&&(c_{u,1})^{\dagger}&&...&&(b_{u,M})^{\dagger} \end{pmatrix}\] is the vector of all electron creation operators in unit cell $u$, as indicated in Fig. \ref{stripfig}. The matrices $\mathbf{H}_{u,w}$ describe hopping from unit cell $w$ to unit cell $u$. 

Starting with the Schr\"odinger equation one can find a recursive relation for the wavefunctions as

\begin{align}
\mathbf{M}_u\cdot \begin{pmatrix} \Psi_{u+1}\\ \Psi_{u} \end{pmatrix}=&\begin{pmatrix} \Psi_{u}\\ \Psi_{u-1} \end{pmatrix} ,
\label{eq:grptransfermatrix}
\end{align}
with the transfer matrix
\begin{align}
\mathbf{M}_u= &\begin{pmatrix} 0&\mathbf{I} \\ \mathbf{A}_u & \mathbf{B}_u  \end{pmatrix},
\end{align}
where
\begin{align}
\mathbf{A}_u=&-\left(\mathbf{H}_{u,u-1}\right)^{-1} \mathbf{H}_{u,u+1}, \nonumber \\ \mathbf{B}_u=&\left(\mathbf{H}_{u,u-1}\right)^{-1}\left(E_F\mathbf{I}-\mathbf{H}_{u,u}\right). \nonumber
\end{align}
$\Psi_{u}=\begin{pmatrix}\psi_{u,1}^A&&\psi_{u,1}^B &&\psi_{u,1}^C&&...&&\psi_{u,M}^B\end{pmatrix}^T$ is a vector of the amplitudes on all sites in one unit cell and $E_F$ our chosen Fermi energy. Using appropriate initial wavefunctions and Eq. \eqref{eq:grptransfermatrix}, we can calculate the wavefunction in every unit cell of our system. Our initial condition is an incoming right-moving wave in lead L such that there are only transmitted waves in lead R and the initial and reflected waves in lead L. Once we have the wavefunctions in both leads we can determine the transmission matrix. For computational convenience we calculate backwards through the system starting in lead R. 

The first step is therefore to determine the eigenmodes of lead R which are given by the eigenvalues $\lambda_n$ and eigenvectors $\mathbf{\Psi}_n$ of the transfer matrix $\mathbf{M}_w$ for $w$ in lead R.  For propagating modes $\lambda_n=e^{i k_n a}$  and the eigenvalue is a root of unity, while otherwise the amplitude is increasing or decaying. Depending on the energy level we choose, we get $p$ pairs of propagating modes (left and right moving) with eigenvalues $e^{\pm i k_n a}$ ($a$ is the length scale of our system and will be set to 1) and eigenvectors $\mathbf{\Psi}_{\pm k_n}$ and $q$ pairs of exponentially increasing or decaying modes (resulting from imaginary momentum modes) with eigenvalues $e^{\pm \kappa_n a}$ and eigenvectors $\mathbf{\Psi}_{\pm \kappa_n}$. The physical eigenmodes of lead R are the starting points for Eq. \eqref{eq:grptransfermatrix}. Taking $w$ to be a unit cell in lead R and $u$ a unit cell in lead L, as illustrated in Fig. \ref{stripfig}, we can determine $2p+q$ physical wavefunctions in lead L using
\begin{equation}
\mathbf{\Psi}^L_{k_n}=\begin{pmatrix} \Psi_{u+1} \\ \Psi_{u} \end{pmatrix} =\mathbf{M}_{u+1} \cdot ... \cdot \mathbf{M}_{w-1} \cdot \mathbf{M}_w \cdot \mathbf{\Psi}^R_{k_n}
\label{eq:backprop}
\end{equation}
\noindent and analogously for $\mathbf{\Psi}^L_{\kappa_n}$. These wavefunctions have to be a superposition of all possible eigenvectors of $\mathbf{M}$ in the leads and can be written as
\begin{equation}
\mathbf{\Psi}^L_{k_n}=\sum_{m=1}^{2p}{\alpha_{k_m,k_n} \mathbf{\Psi}^R_{k_m} e^{iNk_m}}+\sum_{m=1}^{2q}{\alpha_{\kappa_m,k_n} \mathbf{\Psi}^R_{\kappa_m} e^{N\kappa_m}}
\label{eq:lincomb}
\end{equation}
and analogously for $\mathbf{\Psi}^L_{\kappa_n}$ with $N=w-u$ and $k_{p+n}=-k_n$, $\kappa_{q+n}=-\kappa_n$ for brevity. $\alpha_{q^{}_m,q'_n} \in \mathbb{C}$ is the coefficient of mode $q_m$ in lead L when mode $q'_n$ was propagated from lead R to lead L. 

Since we know that the eigenvalues of the transfer matrix are of the form $e^{\pm ik}$ and $e^{\pm \kappa}$, the momenta $k_m$ and $\kappa_m$ can be calculated from the eigenvalues of $\mathbf{M}_w$. Eq. \eqref{eq:lincomb} forms a linear system of equations with a unique solution for the variables $\alpha_{q^{}_m,q'_n}$. In general all the coefficients $\alpha_{q^{}_m,q'_n}$ will be nonzero, that includes the coefficients of the unphysical modes. Although we now essentially have corresponding incoming flux amplitudes in the left lead and outgoing flux amplitudes in the right lead, we cannot calculate the transmission coefficients from those because the picture is not physical. The wavefunction in the left lead contains exponentially increasing contributions. Instead of choosing a single outgoing mode in the right lead (which by itself gives us unphysical contributions in the left lead) we take a superposition of physical modes, i.e., propagating and exponentially decaying modes, in the right lead. The coefficients of this superposition have to be chosen such that (after propagating backwards through the strip) the unphysical modes in the left lead cancel each other. However this cancellation only works for a superposition of one propagating mode with all the exponentially decaying modes. This procedure is illustrated in Fig. \ref{fig:schematic2}. 
We are not interested in the left-going modes in the right lead since we are looking at the case of an incoming wave from the left. To cancel the unphysical modes in the left lead we need to solve $p$ systems of equations like (cf. Fig. \ref{fig:schematic2})
\begin{equation}
\begin{pmatrix} \alpha_{\kappa_1,k_m}&\alpha_{\kappa_1,\kappa_1} & \cdots & \alpha_{\kappa_1,\kappa_q} \\  \vdots & \vdots & \vdots & \vdots \\
\alpha_{\kappa_q,k_m}&\alpha_{\kappa_q,\kappa_1} & \cdots & \alpha_{\kappa_q,\kappa_q}  \end{pmatrix} 
\begin{pmatrix} \beta_{k_m,0} \\ \beta_{k_m,1} \\ \vdots \\ \beta_{k_m,q} \end{pmatrix} = \begin{pmatrix} 0\\ \vdots \\0 \end{pmatrix},
\label{eq:solve}
\end{equation}
\noindent one for each $k_m$ corresponding to a mode propagating to the right. The $\beta_{k_m,n}$ are the coefficients of the superposition in lead R that we want to determine. The index $k_m$ denotes which propagating mode is part of the superposition. The index $n$ shows which mode belongs to this coefficient, $0$ for the propagating mode and $n=1...q$ for the $q$ decaying modes. The $\alpha_{q^{}_n,q'_m}$ are calculated from Eq. \eqref{eq:lincomb} and are the coefficients of the unphysical modes that we want to cancel. $\alpha_{\kappa_1,k_m}$ is the coefficient of the (exponentially increasing) mode $\kappa_1$ in the lead L when the (propagating) mode $k_m$ was propagated from lead R backwards through the system. The set of linear equations \eqref{eq:solve} is under determined and we can therefore choose $\beta_{k_m,0}=1$ without loss of generality.
The coefficients $\beta_{k_m,0}$ and $\beta_{k_m,n}$ give us a superposition of physical modes in lead R which is caused by a superposition of only physical modes in lead L. This is illustrated in Fig. \ref{fig:schematic2} where the entire system is now made up out of only physical modes.
In order to determine the transmission coefficients, we are interested in the coefficients $\gamma$ of the propagating modes in lead L. Taking into account the necessary superpositions, the coefficients for the right-moving modes are given by
\begin{equation}
\gamma_{k_n,k_m} = 1\cdot \alpha_{k_n,k_m} + \sum_{l=1}^q{\beta_{k_m,\kappa_l} \alpha_{k_n,\kappa_l}}
\label{eq:gamma}
\end{equation}
$\gamma_{k_n,k_m}$ is the coefficient of mode $k_n$ in the left lead when the outgoing mode in the right lead is $k_m$.
The transmission coefficients can now be calculated by requiring probability current conservation. This leads to the following system of equations:
\begin{equation}
\begin{pmatrix} T^{LR}_{11} & T^{LR}_{12} & \ldots & T^{LR}_{1p} \\ T^{LR}_{21} & \ddots \\ \vdots & & & \vdots \\ T^{LR}_{p1} &  & \ldots & T^{LR}_{pp} \end{pmatrix} \cdot \mathbf{\Gamma} = \mathbf{V},
\label{eq:tmatrix1}
\end{equation}
\noindent with $\mathbf{\Gamma}_{nm}=\gamma_{k_n,k_m} \sqrt{|v_{k_n}|}$ and $\mathbf{V}_{nm}=\sqrt{|v_{k_n}|}\; \delta_{nm}$. $v_{k_n}= \frac{1}{\hbar} \frac{\partial E}{\partial k}|_{k=k_n}$ is the velocity of mode $k_n$. The $T_{nm}$ are the transmission coefficients with $|T^{LR}_{nm}|^2$ the probability that an electron coming in through mode $k_m$ is transmitted through the conductor into mode $k_n$.
\vspace{42pt}

 \bibliographystyle{apsrev}

 \end{document}